# STRENGTHENING OF COPPER BY CARBON NANOTUBES
## Konstantin Borodianskiy

*Department of Chemical Engineering, Biotechnology and Materials, Ariel University, Ariel 40700, Israel*
*konstantinb@ariel.ac.il*

The influence of a modifier based on multi walled carbon nanotubes (MWCNT) is investigated using C11000 copper alloy. The influence of the modifier addition into the melt was investigated using tensile test, hardness measurements, X-ray diffraction method and microstructural investigations. It was evaluated that the yield and tensile strengths of the metal increased due to the microstructural changes in the formed metal after addition of 0.01 wt. % of the MWCNTs. It was also evaluated that the addition of mentioned amount of MWCNT into alloy has no influence to the phase composition of the formed metal.

**Introduction**

The majority of research works in the field of metallurgy deals with obtaining of the metals with required properties. Usually, advanced properties can be reached by traditional alloying process, where some additives added to the formed material. Unfortunately, the addition of those materials is relatively expensive process.

Copper is extensively applied in different applications because of its advanced properties such as high ductility, high hardness, and high electrical conductivity. One of the main challenges of today's copper-producing industry is to obtain high strength metal. Usually, Cu can be strengthened by solid solution mechanism by the addition of some additives [1] or by cold work processing [2-3]. Nowadays, nanotechnology is also involved into the copper production. Majority of those works describe so called powder metallurgy method of nanomaterials incorporation into Cu [4-5]. That will be logical to assume that copper strengthening behaves similar as aluminum due to their ductile properties. As for aluminum, our previous works described the technology of nanocompounds addition into Al alloys [5-9].

A wide research has been done for the past two decades in synthesis and properties investigation of multi walled carbon nanotubes (MWCNT) [10-13]. MWCNTs have remarkable physical and mechanical properties such as electrical and thermal conductivity, high Young's modulus and high flexibility. Those investigations showed that MWCNTs have pronounced advanced mechanical properties. Therefore, their incorporation into the metallic structure should lead to the strengthening of the bulk metal.

The goal of the presented workis to perform the modification of electrolytic C11000 alloy by the addition of multi walled carbon nanotubes (MWCNT).

**Experimental**

2 kg of copper C11000 ingots werecharged into graphite crucible and melted in a lab furnace Top 16/R up to 1150 ºC.The melt was coated by protective flux and the melt was subjected to a standard refinement procedure.

MWCNT with an outer diameter of 30-50 nm, length of 10-20 μm and purity of > 95 wt.% were used in the work. First, they were mechanochemically treated with copper powder (99+, 325Mesh) in planetary ball mill PM 100 (Retsch GmbH) for 10 min. The obtained mixture undergoes a standard pressing procedure. The produced modifier contained 0.01 wt. % of MWCNT was added into the molten Cu. Then, the melt was poured into a preheated up to 200°C permanent grey cast iron mold. The obtained poured specimens dimensions illustrated in Fig. 1.

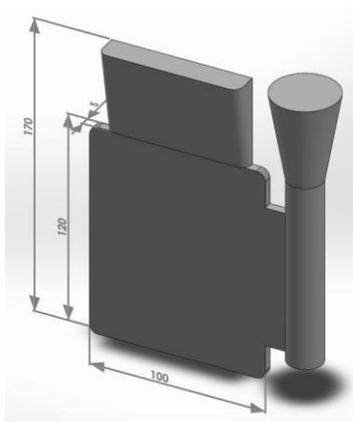

**Fig. 1.** Schematics of poured specimen.

3 samples were cut from the poured specimen, machined to the required dimensions (Fig. 2) and tested using Lloyd EZ50 universal testing machine according to ASTM B 108-01.

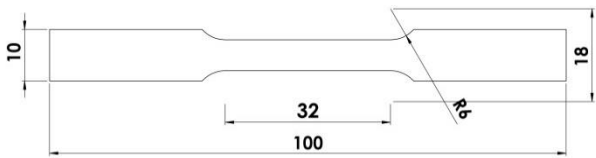

**Fig. 2.** Schematics of the sample subjected to mechanical properties tests.

Hardness measurements were conducted using Vickers Hardness Tester FV-810. The hardness was determined as the mean of five measurements for each sample under a load of 10 kg for 10 s.

Phase analysis evaluated by Panalytical X'Pert Pro X-ray powder diffractometer at 40kV and 40mA. The patterns were measured and recorded at the 2Θ range from 40° to 100° (step size/time per step: 0.02°/2s).

Microscopic examinations were carried out using Olympus BX53MRF-S optical microscope and analyzed by Clemex image analysis software.

**Results and Discussion**

The obtained tensile strength results of modified and unmodified alloy presented in Fig. 3. Tensile test results showed that copper tensile and yield strengths are improved by 30 % after the addition of MWCNT into the melt. Simultaneously, we recognized a slight reduction in the metals elongation.

Hardness test was conducted and its results are presented in Fig. 4.

The hardness test evaluated enhancement in Vickers hardness of the modified alloy more than twice. This behavior can be associated with the presence of the MWCNT into the alloy. Based on the results, it can be derived that the added MWCNT were homogeneously distributed into the alloy because the hardness measurements were detected on a big area of the alloy.

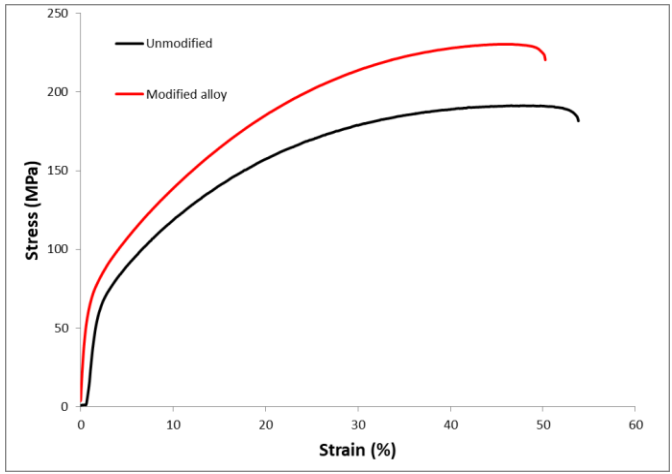

**Fig. 3.** Stress-strain graph performed for unmodified Cu alloy and Cu alloys subjected to modification by MWCNT.

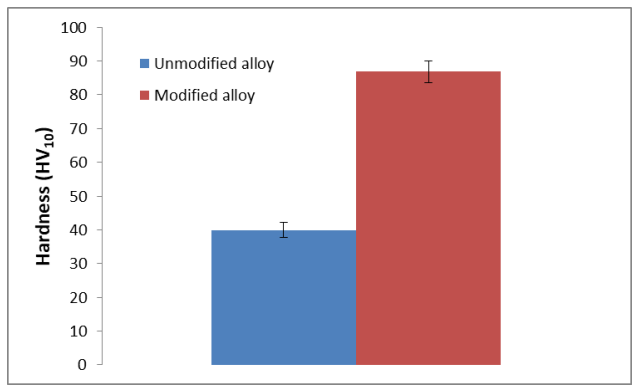

**Fig. 4.** Hardness for unmodified Cu alloy and Cu alloy subjected to modification by MWCNT.

It is also logically to evaluate any detectable changes in phase composition of the modified alloy as the result of the addition of MWCNTs. Therefore, XRD investigation was applied and the obtained patterns presented in Fig. 5.

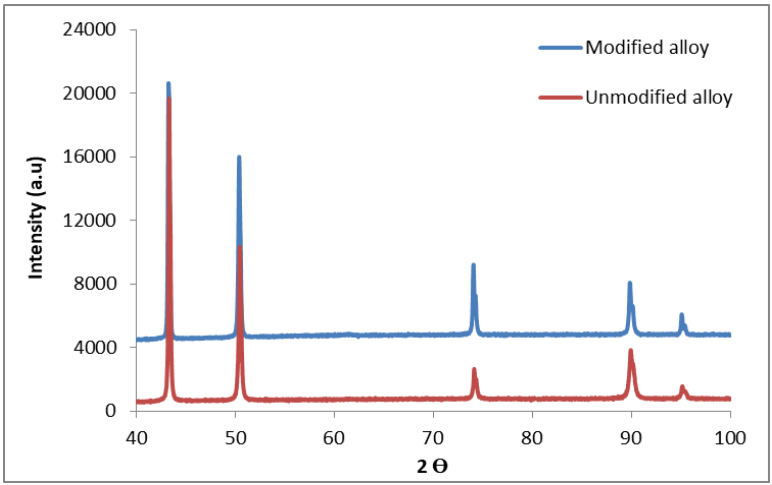

**Fig. 5.** XRD patterns of the unmodified Cu alloy and Cu alloy subjected to modification by MWCNT.

It is evident from the XRD results that the obtained XRD patterns are completely the same e.g., there no new phase was formed during the modification process. In other words, the additives have no influence to the materials phase composition.

Microstructural changes of the modified alloy compared to the unmodified are presented in Fig. 6.

It is clearly seen on the presented micrographs, the unmodified microstructure consists of a coarse, non-oriented and non-homogeneous grains. After the modification process, the grains became more columnar and more homogeneously distributed.

Based on the obtained microstructures, the average grain size was calculated and the obtained results presented in Table 1.

As expected, the average grain size of the modified alloy was reduced compared to the unmodifiedthatis well correlated with the strengthening theory.According to the theory, the metallic alloy can be strengthened as its microstructure becomes finer because of the high density of the formed grain boundaries which are blocking the dislocation movement.

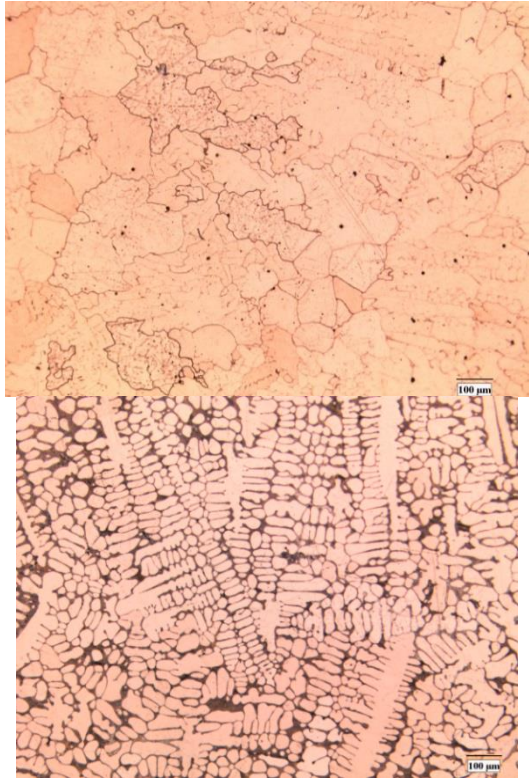

**Fig.6.** Microstructures of unmodified Cu alloy (left) and Cu alloy subjected to modification by MWCNT (right).

Table 1

Average grain size calculations of the unmodified Cu alloy and Cu alloy subjected to modification by MWCNT

|  | Grain average length [μm] | $D_{50}$ [μm] | $D_{90}$ [μm] |
|---|---|---|---|
| Unmodified alloy | 103.12±35.61 | 97.59 | 157.59 |
| Modified alloy | 73.71±30.42 | 69.06 | 112.55 |

## Conclusions

The following can be concluded from the work:

1. Copper can be subjected to modification process by MWCNTs.
2. Tensile test and hardness measurements shows that the modified copper has an advanced properties compared to those of non-modified.
3. Addition of MWCNTs into copper has no influence to the phase composition of the formed metal.

**Acknowledgements**

The current study was supported by the Chief Scientist of the Israel Ministry of Economy and Industry (The Kamin Program Grant No. 56268).